\documentstyle[12pt]{article}
\def \ie {i.~e.~} 
\def\BT{B\"acklund transformation}
\def\CRAS{C.~R.~Acad.~Sc.~Paris}
\def \Log {\mathop{\rm Log}\nolimits}
\def \sinh{\mathop{\rm sinh}\nolimits}
\def \sech{\mathop{\rm sech}\nolimits}
\def \bfY {{\bf Y}}
\begin{document}

\pagestyle{plain} % page numbering

\vskip 0.3 truecm

\begin{center}
 {\bf Various truncations in Painlev\'e analysis of PDEs}
\end{center}

\vskip 0.5 truecm

{\bf R.~Conte}

\medskip
Service de physique de l'\'etat condens\'e,
CEA Saclay,
\hfill \break \indent
F--91191 Gif-sur-Yvette Cedex,
France
\medskip

\baselineskip=12truept
\vskip 0.8 truecm

{\it Abstract}.
The ``truncation procedure'' initiated by Weiss {\it et al.}~is best
understood as a Darboux transformation.
If it leads to the Lax pair of the PDE under study,
the B\"acklund transformation follows by an elimination,
thus proving the integrability.
We present the state of the art of this powerful technique.
The easy situations were all handled by the WTC one--family truncation
and its homographically invariant version.
An updated version of this method has been recently developed,
which is now able to handle the Kaup-Kupershmidt and Tzitz\'eica equations.
It incorporates a new feature,
namely the distinction between two entire functions usually mingled,
which are shown to be linked by formulae established
by Gambier for his classification.

\vfill
Nonlinear dynamics~: integrability and chaos, Tiruchirapalli, 12--16 Feb 1998,

ed.~S.~Daniel.
% World Scientific? Springer?
\hfill 
30~July~1998
\hskip 1.0 truecm
S98/047
\hskip 1.0 truecm
solv-int/9812008
\eject

\tableofcontents 
\vfill \eject

% ==========================================================================
\section{Statement of the general problem}
\indent

The problem we address is the following.
When a partial differential equation (PDE) has some good reasons to be
``integrable'',
please find its \BT\ (BT) \cite{Backlund1883,RogersShadwick},
and do it by singularity analysis {\it only}.

There are two reasons why one should tackle such a problem.
The first one is the success of Weiss \cite{Weiss1983} to solve it
in an algorithmic way for several well known PDEs.
The second one is just our intimate conviction that
a well conducted singularity analysis should capture the global
features of any PDE.

The PDEs which we consider are those which satisfy the necessary conditions 
(i.e.~which ``pass the Painlev\'e test'' \cite{WTC,Cargese96Musette})
for the absence of movable critical singularities in their general solution
(the ``Painlev\'e property'' (PP)).
Then,
in order to prove the sufficiency of these conditions,
the next step is to find a BT since the existence of a BT is often taken as
a definition of the word ``integrability''.

Let us first recall what a \BT\ is.
A BT between two given PDEs 
\begin{equation}
 E_1(u,x,t)=0,\ E_2(U,X,T)=0
%\label{eqNAME}
\end{equation}
is by definition 
(\cite{DarbouxSurfaces} vol.~III chap.~XII, \cite{MatveevSalle})
a pair of relations
\begin{equation}
 F_j(u,x,t,U,X,T)=0,\ j=1,2
%\label{eqNAME}
\end{equation}
with some transformation between $(x,t)$ and $(X,T)$,
in which $F_j$ depends on the derivatives of $u(x,t)$ and $U(X,T)$,
such that the elimination of $u$ (resp.~$U$) between $(F_1,F_2)$
implies
$E_2(U,X,T)=0$ (resp.~$E_1(u,x,t)=0$).
In case the two PDEs are the same, the BT is called the auto-BT.

% ==========================================================================
\section{Transposition to PDEs of the ideas for ODEs}
\label{sectionTransposition}
\indent

For the six ordinary differential equations (ODE) (P1)--(P6) 
which bear his name,
Painlev\'e proved the PP by showing \cite{PaiActa,PaiCRAS1906}
the existence of one (case of (P1)) or two ((P2)--(P6)) 
% nearly entire 
function(s) $\tau=\tau_1, \tau_2$
linked to the general solution $u$ by logarithmic derivatives
\begin{eqnarray}
\hbox{(P1)}:\
& u= & {\cal D}_1 \Log \tau 
\label{eqlogPone}
\\
(\hbox{P}n),\ n=2,\dots,6 :\
& u= & {\cal D}_n (\Log \tau_1 - \Log \tau_2)
\label{eqlogPtwo}
\end{eqnarray}
where the operators ${\cal D}_n$ are linear:
\begin{eqnarray}
& &
{\cal D}_1= - \partial_x^2,\
{\cal D}_2={\cal D}_4=\pm \partial_x,\
{\cal D}_3=\pm e^{-x} \partial_x,\
\\
& &
{\cal D}_5=\pm x e^{-x} (2 \alpha)^{-1/2} \partial_x,\
{\cal D}_6=\pm x (x-1) e^{-x} (2 \alpha)^{-1/2} \partial_x.
\end{eqnarray}
These functions $\tau_1,\tau_2$ have the same kind of singularities than
solutions of {\it linear} ODEs, namely:
they are entire functions for (P1)--(P5), 
and their only singularities for (P6) are three fixed critical points.

For PDEs, similar ideas prevail.
The analogue of (\ref{eqlogPone})--(\ref{eqlogPtwo}) is now the
{\it Darboux transformation}
\begin{equation}
\hbox{DT}:\ 
u = \sum_f {\cal D}_f \Log \tau_f + U
\label{eqDTGeneral}
\end{equation}
linking two different solutions $(u,U)$ of the PDE {\it via} as many 
logarithmic derivatives of ``entire'' functions $\tau_f$ 
as families $f$ of movable singularities.
The linear operators ${\cal D}_f$ are easy to derive from the Painlev\'e test.

A {\it scalar Lax pair} is a system of two linear scalar PDEs
\begin{equation}
\hbox{Lax}:\ 
L_1(U,\lambda) \psi =0,\
L_2(U,\lambda) \psi =0,
\label{eqLax}
\end{equation}
with coefficients depending on the second solution $U$ and,
in the $1+1$-dimensional case, on an arbitrary constant $\lambda$,
which has the property that the vanishing of the commutator $[L_1,L_2]$ is
equivalent to the vanishing of the PDE $E(U)=0$.

Finally, there exists a link
\begin{eqnarray}
& & \forall f\ {\cal D}_f \Log \tau_f = F_f(\psi),
\end{eqnarray}
which most often is the identity $\tau=\psi$,
between the entire functions $\tau_f$ and the entire function $\psi$.

As to the {\it auto-B\"acklund transformation}
\begin{equation}
\hbox{BT}:\ 
F_1(u,U,\lambda)=0,\
F_2(u,U,\lambda)=0
\label{eqBT}
\end{equation}
it is made of two equations resulting from the elimination of $\psi$ 
between the DT (\ref{eqDTGeneral}) and the scalar Lax pair (\ref{eqLax}).

Up to now,
there seem to exist two and only two classes of integrable $1+1$-dimensional
PDEs~:
those who have only one family of movable singularities,
and those who have only pairs of families with opposite principal parts,
similarly to the distinction between (P1) on one side
and (P2)--(P6) on the other side.
Among the $1+1$-dimensional equations,
those with one family include 
KdV, 
the AKNS and the Boussinesq equations;
they also include
the Sawada-Kotera, 
the Kaup-Kupershmidt and the Tzitz\'eica equations
because only one of their two families is relevant
\cite{MC1998,CMG1998}.
Equations with pairs of opposite families
include sine-Gordon, mKdV and Broer-Kaup (two families each),
NLS (four families).

% ==========================================================================
\section{The truncation method for a one-family equation}
\label{sectionTruncationOneFamily}
\indent

Its historical version is the famous WTC truncation method.
We summarize here the most recent state of this widely used method,
now able to handle the Kaup-Kupershmidt and Tzitz\'eica equations.
As compared with the detailed exposition of Ref.~\cite{MC1998},
we remove here the restriction $E(U)=0$, see first step below.

Consider a PDE with only one family of movable singularities
\begin{eqnarray}
& & E(u)=0,
\label{eqPDE}
\end{eqnarray}
and denote ${\cal D}$ the {\it singular part operator} of its unique family.

{\it First step}.
Assume a Darboux transformation (DT) \cite{Darboux1882} defined as
\begin{eqnarray}
& &
u=U + {\cal D} \Log \tau,\
E(u)=0,
\label{eqDT}
\end{eqnarray}
with $u$ a solution of the PDE under consideration,
$U$ an unspecified field which most of the time will be found to be
a second solution of the PDE,
$\tau$ some ``entire'' function.
A consequence of this assumption is the existence of the involution
\begin{eqnarray}
& &
(u,U,\tau) \to (U,u,\tau^{-1}),
\end{eqnarray}
since the operator ${\cal D}$ is linear.

{\it Second step}.
Choose the order two, then three, then \dots,
for the unknown {\it scalar} Lax pair.
The reason why the Lax pair should be defined in scalar, not matrix, form
will become clear at the third step.
Such a second-order scalar Lax pair in canonical form,
written here in the case of two independent variables $(x,t)$,
is
\begin{eqnarray}
L_1 \psi & \equiv & 
\psi_{xx} + {S \over 2} \psi=0,
\label{eqLaxScalar2x}
\\
L_2 \psi & \equiv & 
\psi_{t} + C \psi_{x} - {C_x \over 2} \psi=0,
\label{eqLaxScalar2t}
\\
2 [L_1,L_2] & \equiv & X = S_t + C_{xxx} + C S_x + 2 C_x S=0.
\label{eqCrossSC}
\end{eqnarray}
A third-order scalar Lax pair in canonical form (no $\psi_{xx}$ in $L_1$)
is
\begin{eqnarray}
L_1 \psi & \equiv &
\psi_{xxx} - a \psi_x - b \psi=0,
\label{eqLaxScalar3x}
\\
L_2 \psi & \equiv &
\psi_t  - c\psi_{xx} - d \psi_x - e \psi=0,
\label{eqLaxScalar3t}
\\
{} [L_1,L_2] & \equiv &
X_0 + X_1 \partial_x + X_2 \partial_{x}^2,
\\
X_0  & \equiv &
 - b_t - a e_x + e_{xxx} + b_{xx} c 
 + 3 b c_{xx} + 3 b_x c_x + 3 b d_x + b_x d = 0,
\label{eqX0}
\\ X_1  & \equiv &
 - a_t + 3 e_{xx} + 2 b_xc + a_{xx} c + d_{xxx} + 3 a c_{xx}+ 2 a d_x
\nonumber
\\ & & + 3 a_x c_x + 3 b c_x + a_xd = 0,
\label{eqX1}
\\ X_2  & \equiv &
 (2 a c + c_{xx} + 3 d_x + 3 e)_x=0.
\label{eqX2}
\end{eqnarray}

{\it Third step}.
Choose an explicit link 
\begin{eqnarray}
& &
{\cal D} \Log \tau=f(\psi),
\label{eqTauPsiLink}
\end{eqnarray}
between the function $\tau$
and the solution $\psi$ of a scalar Lax pair.
It will be shown in section \ref{sectionGambier} that,
at each scattering order,
there exists only a finite number of choices (\ref{eqTauPsiLink}),
among them the most frequent one
\begin{eqnarray}
& &
\tau=\psi.
\label{eqTauPsiLink1}
\end{eqnarray}

{\it Fourth step}.
Define the ``truncation'' and solve it,
that is to say~:
with the assumptions 
(\ref{eqDT}) for a DT,
(\ref{eqTauPsiLink}) for a link between $\tau$ and $\psi$,
(\ref{eqLaxScalar2x})--(\ref{eqLaxScalar2t})
or
(\ref{eqLaxScalar3x})--(\ref{eqLaxScalar3t})
for the scalar Lax pair in $\psi$,
%(\ref{eqChix})--(\ref{eqChit}) or 
%(\ref{eqProjRiccatiY1x})--(\ref{eqProjRiccatiY2t})
%for the gradient of $\bfY$,
express $E(u)$ as a polynomial in the derivatives of $\psi$
which is irreducible {\it modulo} the scalar Lax pair.
For the above pairs,
this amounts to eliminate any derivative of $\psi$ of order in $(x,t)$ 
higher than or equal 
   to $(2,0)$ or $(0,1)$ (second order case)
or to $(3,0)$ or $(0,1)$ (third order),
thus resulting in a polynomial of one variable $\psi_x / \psi$ (second order)
or two variables $\psi_x / \psi, \psi_{xx} / \psi$ (third order)
\begin{eqnarray}
& &
E(u)= \sum_{j=0}^{-q} E_j(S,C,U) (\psi / \psi_x)^{j+q}
\hbox{ (for second order)},
\label{eqTruncationOrder2}
\\
& &
E(u)= \sum_{k \ge 0} \sum_{l \ge 0} E_{k,l}(a,b,c,d,e,U) 
(\psi_x / \psi)^k (\psi_{xx} / \psi)^l
\hbox{ (for third order)}.
\label{eqTruncationOrder3}
\end{eqnarray}
Finally,
solve the set of {\it determining equations}
\begin{eqnarray}
\forall j\
& &
E_{j}(S,C,U) =0
\hbox{ (second order)}
\label{eqDetermining2}
\\
\forall k\
\forall l\
& &
E_{k,l}(a,b,c,d,e,U) =0
\hbox{ (third order)}
\label{eqDetermining3}
\end{eqnarray}
for the unknown coefficients $(S,C)$ or $(a,b,c,d,e)$ as functions of $U$,
and find the PDE which $U$ satisfies.
If this PDE is the same as (\ref{eqPDE}),
one is on the way to an auto-BT,
otherwise to an hetero-BT.

The second, third and fourth steps must be repeated until a success occurs.
The process is successful if and only if all the following conditions are met
\begin{enumerate}
\item
$U$ comes out unconstrained, apart from being a solution of some PDE,
\item
(if an auto-BT is desired) the PDE satisfied by $U$ is identical to 
(\ref{eqPDE}),
\item
the vanishing of the commutator $[L_1,L_2]$ is equivalent to the vanishing
of the PDE satisfied by $U$,
\item
in the $1+1-$dimensional case only
and if the PDE satisfied by $U$ is identical to (\ref{eqPDE}),
the coefficients depend on an arbitrary constant $\lambda$,
the spectral or B\"acklund parameter.
\end{enumerate}

At this stage, one has obtained the DT and the Lax pair.

{\it Fifth step}.
Obtain the two equations for the BT by eliminating $\psi$ \cite{Chen1974}
between the DT and the scalar Lax pair.
This sometimes uneasy operation when the order $n$ of the scalar Lax pair
is too high
may become elementary by introducing a $(n-1)$-component pseudopotential $\bfY$
in place of $\psi$
and by eliminating the appropriate components of $\bfY$ rather than $\psi$.
Assume for instance that $\tau=\psi$ and ${\cal D}=\partial_x$.
Then Eq.~(\ref{eqDT}) reads
\begin{eqnarray}
& &
Y_1=u-U
\label{eqY1uU}
\end{eqnarray}
and the BT is computed as follows~:
eliminate all the components of ${\bf Y}$ but $Y_1$ between the equations for 
the gradient of ${\bf Y}$,
then in the resulting equations substitute $Y_1$ as defined in (\ref{eqY1uU}).

For the scalar second-order Lax pair 
(\ref{eqLaxScalar2x})--(\ref{eqLaxScalar2t}),
the equivalent one-component Riccati system for
$\bfY=\psi_x / \psi = \chi^{-1}$ is
\begin{eqnarray}
& & (\chi^{-1})_x= - \chi^{-2} - {S \over 2},
%& & \chi_x= 1 + {S \over 2} \chi^2
\label{eqChix}
\\
& & (\chi^{-1})_t=  C \chi^{-2} - C_x \chi^{-1} + {C S + C_{xx} \over 2}.
%& & \chi_t= -C + C_x \chi -{C S + C_{xx} \over 2} \chi^2.
\label{eqChit}
\end{eqnarray}
For the scalar third-order Lax pair 
(\ref{eqLaxScalar3x})--(\ref{eqLaxScalar3t}),
an equivalent two-component pseudopotential
is the projective Riccati one
$\bfY=(Y_1,Y_2)$
\cite{MC1991,MCGallipoli1991}
\begin{eqnarray}
Y_1 & = & {\psi_{x}  \over \psi},\
Y_2={\psi_{xx} \over \psi},\
\\
Y_{1,x} & = & - Y_1^2 + Y_2,
\label{eqProjRiccatiY1x}
\\ 
Y_{2,x} & = & -Y_1 Y_2 + a Y_1 + b,
\label{eqProjRiccatiY2x}
\\
Y_{1,t} & = & - (d Y_1 + c Y_2) Y_1 + (a c+d_x)Y_1 + (c_x+d)Y_2 +e_x+b c
\label{eqProjRiccatiY1t}
\\ 
        & = & (c Y_2 + d Y_1 + e)_x,
\label{eqProjRiccatiY1tCons}
\\ 
Y_{2,t} & = & - (d Y_1 + c Y_2) Y_2 +(2 a c_x+a_x c+b c+d_{xx}+a d+2e_x)Y_1
\nonumber
\\ & &
 +(c_{xx}+2 d_x+a c) Y_2 +2 b c_x+b_x c+b d+e_{xx},
\label{eqProjRiccatiY2t}
\\
(Y_{1,t})_x-(Y_{1,x})_t & = &  X_1 + X_2 Y_1,
\\
(Y_{2,t})_x-(Y_{2,x})_t & = & -X_0 + X_2 Y_2.
\end{eqnarray}
If the computation of the BT requires the elimination of $Y_2$,
this BT is 
\begin{eqnarray}
& &
Y_{1,xx} + 3 Y_1 Y_{1,x} + Y_1^3 - a Y_1 - b =0,
\label{eqBTxOrder3}
\\
& &
Y_{1,t} - (c Y_{1,x} + c Y_1^2 + d Y_1 + e)_x=0,
\label{eqBTtOrder3}
\\
& &
(Y_{1,xx})_t-(Y_{1,t})_{xx}=X_0 + X_1 Y_1 + X_2 Y_1^2=0,
\end{eqnarray}
in which $Y_1$ is replaced by an expression of $u-U$, e.g.~(\ref{eqY1uU}).

To fix the ideas, let us give an example \cite{MusetteSainteAdele}.
The AKNS equation \cite{AKNS}
\begin{eqnarray}
& &
E(u) \equiv u_{xxxt} 
+ 4 \alpha^{-1} (2 (u_x - \beta) u_{xt} + (u_t - \gamma) u_{xx})= 0
\end{eqnarray}
admits the single family
\begin{eqnarray}
& &
u \sim \alpha \chi^p,\
p=-1,\
q=-5,\
\hbox{indices } (-1,4,6),\
{\cal D}= \alpha \partial_x,
\end{eqnarray}
so the assumption for the DT is
\begin{eqnarray}
& &
u=U + {\cal D} \Log \tau,\
E(u)=0,
\end{eqnarray}
Let us choose,
at second step the scalar Lax pair 
(\ref{eqLaxScalar2x})--(\ref{eqLaxScalar2t}) for $\psi$,
at third step the link (\ref{eqTauPsiLink1}) between $\tau$ and $\psi$.
Then there are only three non identically zero determining equations 
(\ref{eqDetermining2})
\begin{eqnarray}
 E_2 & \equiv & 4 \alpha S C + 8 (U_{t} - \gamma) - 16 C (U_{x} - \beta) = 0,
\\
 E_3 & \equiv & - \alpha (C S_x + 4 S C_x) + 16 C_x (U_{x} - \beta) 
- 8 U_{xt} + 4 C U_{xx} = 0,
\\
 E_5 & \equiv & E(u_1) + (\alpha/2) (2 S S_t - C S S_x - S_{xxt} - S_x C_{xx})
\nonumber
\\
& &
- 2 S_x (U_{t} - \gamma) - 4 S_t (U_{x} - \beta) - 4 S U_{xt} 
+ 2 (S C + C_{xx}) U_{xx} = 0.
\end{eqnarray}
Their solution for $(S,C)$ depends on an arbitrary complex constant parameter
$\lambda$
and the only constraint on the field $U$ is that it must satisfy the AKNS PDE
\begin{eqnarray}
& &
S=(4/ \alpha) (U_{x} - \beta) - 2 \lambda,\ 
\\
& &
C = (U_{t} - \gamma)/(\alpha \lambda),\
\\
& &
X \equiv E(U)/(\alpha \lambda)=0.
\end{eqnarray}
The BT is the result of the substitution $\chi^{-1}=(u-U)/ \alpha$ in 
(\ref{eqChix})--(\ref{eqChit}).

% ==========================================================================
\section{Two common errors in the one-family truncation}
\indent

Before proceeding, it is worth warning the reader against two errors
frequently made in the method of section \ref{sectionTruncationOneFamily}.

% ==========================================================================
\subsection{The constant level term is not another solution}
\indent

Consider the one-family truncation as done by WTC
(the subscript $T$ means ``truncated'')
\begin{equation}
u_T^{\rm WTC}=\sum_{j=0}^{-p} u_j^{\rm WTC} \varphi^{j+p}
\label{eqWeissTruncationu}
\end{equation}
in which $\varphi$ is the function defining the singularity manifold.

In the WTC truncation,
one considers three solutions of the PDE
\begin{enumerate}
\item
the lhs $u_T^{\rm WTC}$ of the truncation (\ref{eqWeissTruncationu}),

\item
the ``constant level'' coefficient $u_{-p}^{\rm WTC}$,

\item
the field $U$ which appears in the Lax pair after the successful completion
of the method.

\end{enumerate}

The frequently encountered argument
``The constant level coefficient $u_{-p}^{\rm WTC}$ also satisfies the PDE,
therefore one has obtained a BT''
is wrong.
This is obvious, since nonintegrable PDEs, which have no BT,
nevertheless have this property.
One can check it by taking the explicit example of a nonintegrable
PDE \cite{CM1989}.

A hint that the above argument might be wrong is the fact,
observed on all successful truncations,
that the $U$ in the Lax pair is {\it never} $u_{-p}^{\rm WTC}$.
Let us prove this fact, with the homographically invariant analysis 
\cite{Conte1989}.
The truncation of the same variable in the invariant formalism is
\begin{equation}
u_T=\sum_{j=0}^{-p} u_j \chi^{j+p},
\label{eqInvarTruncationu}
\end{equation}
in which $\chi$ is given by
\begin{eqnarray}
& &
{1 \over \chi}
=
{\varphi_x \over \varphi - \varphi_0} - {\varphi_{xx} \over 2 \varphi_x}.
\label{eqchi}
\end{eqnarray}
This $u_T$ depends on the movable constant $\varphi_0$ and one has
\begin{eqnarray}
& &
\left\lbrace
\begin{array}{ll}
\displaystyle{
u_T^{\rm WTC} = u_T (\varphi_0=0)
}
\\
\displaystyle{
u_{-p}^{\rm WTC} = u_T (\varphi_0=\infty).
}
\end{array}
\right.
\end{eqnarray}
Since the results of the truncation do not depend on the movable constant
$\varphi_0$,
this proves that the lhs $u_T^{\rm WTC}$ of the truncation
and the constant level coefficient $u_{-p}^{\rm WTC}$
are not considered as distinct by the truncation procedure.
Since the $U$ in the Lax pair cannot be the truncated $u$
(otherwise one would not have a Darboux transformation),
this ends the proof.

% ==========================================================================
\subsection{The WTC truncation is suitable iff the Lax order is two}
\indent

We mean the truncation as originally introduced,
not its updated version of section \ref{sectionTruncationOneFamily}.

When the Lax pair has second order, everything is consistent.
When the Lax pair has a higher order, e.g.~three,
the original method,
as well as its original invariant version \cite{MC1991},
presents the following inconsistency.
In a first stage,
it generates the $-q+p$ equations $E_j(S,C,U)=0$
of formula (\ref{eqTruncationOrder2}),
which intrinsically correspond to a {\it second}-order scattering problem
(and this is precisely the inconsistency),
and in a second stage it injects in each of these $-q+p$ equations
a link between $(S,C)$ and the scalar field $\psi$ of the Lax pair of
{\it higher} order,
thus generating determining equations which are hybrid between the second order
and the higher one.
The first nearly correct treatment seems to have been made in 
ref.~\cite{MCGallipoli1991}.

For the same reason, 
in order to obtain the Lax pair when its order is higher than two,
it is also inconsistent to consider the so-called
{\it singular manifold equation} (SME) \cite{WTC,Conte1989,Pickering1996}.
The SME is by definition the relation between $S$ and $C$ obtained by 
elimination of $U$ between their expression.
For instance, in the above AKNS equation, this is
\begin{eqnarray}
& &
{S_t \over C_x} - 4 \lambda=0.
\end{eqnarray}
When the Lax order is three,
the correct extension of this notion would be the set of
three relations on $(a,b,c,d)$ resulting from the 
elimination of $U$ between the four coefficients of the Lax pair
($e$ is derivable from (\ref{eqX2}) so we discard it),
but this seems of little interest.

Although these inconsistencies may still provide the full result
for some ``robust'' equations 
(Boussinesq \cite{Weiss1985Bq}, 
 Sawada-Kotera \cite{Weiss1984KKSK},
 Hirota-Satsuma \cite{MC1991}),
there do exist equations for which it leads to a failure,
and the Kaup-Kupershmidt equation \cite{MC1998} is one of them.

Even when the Lax order is higher than two,
the assumption of a second-order Lax pair may lead to an interesting result,
such as a Miura transformation to another PDE.
This is the case for the Kaup-Kupershmidt equation
\begin{eqnarray}
{\rm KK}(u)
& \equiv &
\beta u_t + \left(
u_{xxxx} + {30 \over \alpha} u u_{xx} + {45 \over 2 \alpha} u_x^2
 + {60 \over \alpha^2} u^3 \right)_x = 0,
\label{eqKKcons}
\end{eqnarray}
whose family $u \sim -(\alpha/2) \chi^{-2}$ generates only one nonzero
equation (\ref{eqDetermining2})
\begin{eqnarray}
& & E_4 / \alpha \equiv S_{xx} + S^2/4 - \beta C = 0,
\end{eqnarray}
which is therefore the SME.
The elimination of $C$ with the equation $X=0$ (\ref{eqCrossSC})
implies that $(\alpha/12) S$ satisfies the Sawada-Kotera equation 
\cite{Weiss1984KKSK}
\begin{eqnarray}
{\rm SK}(w)
& \equiv &
\beta w_t 
+ \left(w_{xxxx}+{30 \over \alpha} w w_{xx}+{60 \over \alpha^2} w^3\right)_x=0.
\label{eqSKcons}
\end{eqnarray}
Together with the truncated series for $u$
\begin{eqnarray}
& & u= - (\alpha/2) \chi^{-2} - (\alpha/6) S,
\end{eqnarray}
this provides the link between the KK and the SK equations \cite{FG1980}.

% ==========================================================================
\section{The help of Gambier for the BT of a PDE}
\label{sectionGambier}
\indent

Let us show that at each scattering order
there exists only a finite number of choices (\ref{eqTauPsiLink}).

One of the two PDEs defining the BT to be found can be made an ODE,
e.g.~(\ref{eqChix}) or (\ref{eqBTxOrder3}).
This nonlinear ODE for $Y$,
with coefficients depending on $U$ and,
in the $1+1$-dimensional case, on an arbitrary constant $\lambda$,
has two properties \cite{MC1998}.
Firstly, it is linearizable since, by an elimination process \cite{Chen1974},
it results from the Lax pair, a linear system, and the Darboux transformation.
Secondly,
it has the Painlev\'e property since it is linearizable.
Therefore, if its order is small (at most three),
it belongs to the appropriate finite list established by the Painlev\'e
school between 1900 and 1910.

These very special nonlinear ODEs provide a link to both the Lax pair,
{\it via} their linearizing transformation,
and the Darboux transformation,
{\it via} an involution which leaves them invariant.

The only ODE of first order and degree one with the PP is
the Riccati equation,
it is linearizable into a second-order linear equation and this defines a
unique choice (\ref{eqTauPsiLink})
for describing scattering problems which have order two. 

% NLS is excluded (order 2*2=4)
Next, the ODEs of order two and degree one with the PP
which are also linearizable 
bear the numbers (Ref.~\cite{GambierThese} p.~21)
 5, %                  linear = order 3 projective Riccati = Burgers 2
 6, % IP = Riccati
14, % IP = Riccati
24, % Gambier page 28, linear = order 2
25, % Gambier page 28, linear = order 3
27, % page 51. n=2     linear = order 4 conformal Riccati
    %          n=2     linear = order 3 conformal Riccati particular p 27 l -1
    %          n\not=2 linear = order 2 twice (page 53)
in the classification of Gambier 
of fifty equations inequivalent under the homographic group of transformations,
with the respective orders 
for the associated linear equations,
\ie\ in our context for the unknown scattering problem~:
3, 2, 2, 2, 3, 2 (and 4 or 3 for the case $n=2$ in the equation number 27).

The only two generic choices for describing scattering problems of third order
are therefore the two classes of equivalence numbered 5 and 25 by Gambier.
The representative equation of interest in each class of equivalence is
the ``complete equation''
(in the present section, the symbol $'$ means $\partial_x$)
\begin{eqnarray}
&
Y'' + 3 Y Y' + Y^3 +r Y + q=0,
&
\hbox{(G5)}
\nonumber
\\
&
Y''-3 Y'^2/(4 Y) +3 Y Y'/2 + Y^3/4 -q' (Y'+Y^2)/(2q) -r Y -q=0,
&
\hbox{(G25)}
\nonumber
\end{eqnarray}
in which $q$ and $r$ are two arbitrary functions.
These two classes are equivalent under the birational group,
with the explicit transformation between
G5$(y;q,r)$ and G25$(Y;Q,R)$
\begin{eqnarray}
& &
Y={Q \over 2 z' + z^2 -(Q'/Q) z - R},\
z=y+{Q' \over 2 Q},\
2 y= {Y' \over Y} + Y + {Q' \over Q}, 
\\
& &
r=-R+{Q'' \over Q} - {5 Q'^2 \over 4 Q^2},\
q=-{Q \over 2} -{R' \over 2} + {Q''' \over 2 Q} -{7 Q' Q'' \over 4 Q^2}
+{5 Q'^3 \over 4 Q^3}.
\end{eqnarray}

The linearizing transformations are
\begin{eqnarray}
& \hbox{(G5)} &
Y={\tau' \over \tau}={\psi' \over \psi},\
\psi''' + r \psi' + q \psi=0,
\label{eqlinearG5}
\\
& \hbox{(G25)} &
\left\lbrace
\begin{array}{ll}
\displaystyle{Y={\tau' \over \tau}={q \over 2 z' + z^2 -(q'/q) z - r},\
z={\psi' \over \psi},\ }
\\
\displaystyle{
\psi''' -{3 q' \over 2 q} \psi''
-\left(r + {q'' \over q} - {q'^2 \over q^2} \right) \psi'
-\left({r' \over 2} + {q \over 2} - {q' r \over 2 q} \right) \psi 
=0.
}
\end{array}
\right.
\label{eqTauPsiLink3}
\label{eqlinearG25}
\end{eqnarray}

The involutions
\begin{eqnarray}
& \hbox{(G5)} &
(Y,q,r) \to (-Y,-q,r + 6 Y'),
\label{eqInvolutionG5}
\\
& \hbox{(G25)} &
(Y,q,r) \to (-Y,-q,r - 3 Y' -(q'/q) Y)
\label{eqInvolutionG25}
\end{eqnarray}
% The field cannot depend on Y
characterize the Darboux transformation.

In the third step of the method in section \ref{sectionTruncationOneFamily},
if the chosen scattering order is three,
the formula (\ref{eqTauPsiLink}) is necessarily {\it one of the two}
linearization formulae (\ref{eqlinearG5}) and (\ref{eqlinearG25}),
which thus replace the assumption (\ref{eqLaxScalar3x}).
In the fourth step, 
the unknowns $(a,b)$ in the determining equations (\ref{eqDetermining3})
are then replaced by the unknowns $(q,r)$.

Thanks to the choice (G25),
the BT of the Kaup-Kupershmidt equation,
which had resisted all attempts,
could finally been obtained \cite{MC1998}.

% ==========================================================================
\section{Where to truncate, and with which variable?}
\label{sectionWhere}
\indent

In the process of turning a {\it local} information,
namely an infinite Laurent series expansion,
into a {\it global} one as required by any correct definition
of the word ``integrability'',
one must build some {\it finite} expression for $u$ from this series.
This ``truncation'' can be any closed-form resummation of the series,
and the resulting finite expression for $u$ will ultimately represent the
Darboux transformation.
The resummation variable $Y$ must fulfill two conditions
\begin{enumerate}
\item
$Y$ must be a homographic transform of $\varphi - \varphi_0$,
so that the structure of singularities in the $\varphi$ complex plane has
a one-to-one correspondence with that in the $Y$ complex plane;

\item
$Y$ must vanish as $\varphi - \varphi_0$.
\end{enumerate}

The variable $\chi$ of the invariant Painlev\'e analysis, 
Eq.~(\ref{eqchi}),
is only a particular solution of these two conditions.
To avoid any restriction in the quest for the BT,
one must consider its general solution
\cite{MC1991,Pickering1996}
\begin{eqnarray}
& &
Y^{-1}=B(\chi^{-1} + A),\
%Y={\chi \over B (1 + A \chi)},\
A B \not=0.
\label{eqMostGeneralY}
\end{eqnarray}
Since a homography conserves the Riccati nature of an ODE,
$Y$ satisfies a Riccati system,
easily deduced from the canonical one (\ref{eqChix})--(\ref{eqChit}) 
satisfied by $\chi$.

The advantage of $\chi$ or $Y$ over $\varphi - \varphi_0$ is the following.
The gradient of $\chi$ (resp.~$Y$) is a polynomial of degree two in
$\chi$ (resp.~$Y$),
so each derivation of a monomial increases the degree by one,
while the gradient of $\varphi - \varphi_0$ is a polynomial of degree zero
in $\varphi - \varphi_0$,
so each derivation decreases the degree by one.
Consequently,
when one searches for the admissible values of the rank $-p'$ for stopping
the series in $Z$
\begin{eqnarray}
& &
u=\sum_{j=0}^{-p'} u_j Z^{j+p},\
u_0 u_{-p'} \not=0,\
E=\sum_{j=0}^{-q'} E_j Z^{j+q},\
Z= \hbox{either } \varphi - \varphi_0,\ \chi \hbox{ or }Y,
\end{eqnarray}
one finds two solutions and only two \cite{Pickering1993}
\begin{enumerate}
\item
$p'=p,q'=q$,
in which case the three truncations are identical,
since the three sets of equations $E_j=0$ are equivalent,

\item
for $\chi$ and $Y$ only,
$p'=2 p,q'=2 q$,
in which case the two truncations are different
since the two sets of equations $E_j=0$ are inequivalent
(they are equivalent only if $A=0$).

\end{enumerate}
This second truncation is the only way to find $\sech$-type solitary waves
\cite{Pickering1993},
not only $\tanh$-type waves,
as shown by the elementary identities \cite{CM1993}
\begin{equation} 
   \tanh z - {1 \over \tanh z}= -2 i \sech\left[2 z + i {\pi \over 2}\right],\
   \tanh z + {1 \over \tanh z}=  2   \tanh\left[2 z + i {\pi \over 2}\right].
\end{equation}
Its full applicability is presented in section
\ref{sectionTruncationTwoFamilies}.

The reason why the Riccati representation is preferable is because
it allows several linearizations.
Details can be found in Refs.~\cite{MC1994,Pickering1996}.

% ==========================================================================
\section{The truncation method for an equation with opposite families}
\label{sectionTruncationTwoFamilies}
\indent

When the base member of the hierarchy of integrable equations has more than
a single family,
these families usually come by pairs of opposite singular part operators,
just like (P2)--(P6).
Examples are enumerated at the end of section \ref{sectionTransposition}.
Then the sum of the two opposite singular parts
${\cal D} \Log \tau_1 - {\cal D} \Log \tau_2$
only depends on the variable 
\begin{eqnarray}
& &
Y={\tau_1 \over \tau_2}.
\label{eqYtau12}
\end{eqnarray}

The current status of the method \cite{MC1994,Pickering1996},
which we used to call the two-singular manifold method,
is as follows.
Most of the method for one-family equations still applies,
with the difference that it is much more convenient to represent the
Lax pairs in a Riccati form than in a scalar linear form.
Let us restrict here to second-order scattering problems
and to identity links (\ref{eqTauPsiLink1}) between the two $\tau$ and the
two $\psi$ functions.
Then $Y$ satisfies a Riccati system and, as explained in section
\ref{sectionWhere},
its most general expression is given by (\ref{eqMostGeneralY}).

In the first step, $\tau$ is simply replaced by $Y$ 
in the assumption (\ref{eqDT}) for a DT.

In the second step, the scattering problem is represented by the Riccati
system satisfied by $Y$,
whose coefficients depend on $(S,C,A,B)$.

The fourth step contains the main difference.
Rather than truncating $u$ at the level $j=-p$,
one truncates it at the level $j=-2p$ 
\cite{MC1994,Pickering1996},
in order to implement the two movable singularities $\tau_1=0$ and $\tau_2=0$.
So the truncation is
\begin{eqnarray}
& &
u={\cal D} \Log Y + U,\
\\
& &
Y^{-1}=B(\chi^{-1} + A),\
\\
& &
E(u)=\sum_{j=0}^{-2 q} E_j(S,C,A,B,U) Y^{j+q}=0,\
\\
& &
\forall j\
E_{j}(S,C,A,B,U) =0,
\end{eqnarray}
in which nothing is imposed on $U$.

Let us take the example of the sine-Gordon equation 
\begin{eqnarray}
& &
E(u) \equiv u_{xt} + \alpha (e^u - e^{-u})=0,
\end{eqnarray}
whose two opposite families (opposite in the field $u$) are
\begin{eqnarray}
& &
 e^u \sim -(2/ \alpha) \varphi_x \varphi_t (\varphi - \varphi_0)^{-2},\
\hbox{indices } (-1,2),\
{\cal D}=(2/ \alpha) \partial_x \partial_t.
%\label{eqFamilyOne}
\\
& &
 e^{-u} \sim -(2/ \alpha) \varphi_x \varphi_t (\varphi - \varphi_0)^{-2},\
\hbox{indices } (-1,2),\
{\cal D}=(2/ \alpha) \partial_x \partial_t.
%\label{eqFamilyTwoSG}
\end{eqnarray}
The resulting DT assumption
\begin{eqnarray}
& &
e^u - e^{-u} = (2/ \alpha) \partial_x \partial_t \Log Y + \dots,\ E(u)=0
\end{eqnarray}
can be integrated twice due to the special form of the PDE, resulting in
\begin{eqnarray}
& &
u=-2 \Log Y + W,\ E(u)=0,
\end{eqnarray}
in which nothing is imposed on $W$ (we use $W$ to reserve the symbol $U$
for future use).
The five determining equations $E_j=0,j=0,...,4$ are solved as usual
by ascending values of $j$
\begin{eqnarray}
& &
E_0:\
B^2 e^W ={2 \over \alpha} C,
\\
& &
E_1:\
A=-{1 \over 2} (\Log C)_x,\
\\
& &
E_2 \equiv 0,
\\
& &
E_3:\
S=-F(x) + {C_x^2 \over 2 C^2} - {C_{xx} \over C},
\\
& &
E_4:\
C C_{xt} - C_x C_t + F(x) C^3 - \alpha^2 F(x)^{-1} C=0,
\\
& &
X:\
\alpha F'(x)=0.
\end{eqnarray}
in which $F$ is a function of integration.
So $F(x)$ must be a constant
\begin{eqnarray}
& &
F(x)=2 \lambda^2,
\end{eqnarray}
and $\Log C$ is proportional to a second solution $U$ of the PDE
\begin{eqnarray}
& &
C={\alpha \over 2} \lambda^{-2} e^{U},\
E(U)=0,
\end{eqnarray}
and one has obtained the Darboux transformation 
\begin{eqnarray}
& &
u = -2 \Log y + U,\
y=\lambda B Y,
\label{eqDTLiouvilleandSG}
\end{eqnarray}
in which $y$ satisfies the Riccati system 
\begin{eqnarray}
& &
y_x = \lambda +  U_x y - \lambda y^{2},
\label{eqRiccatiZx}
\\
& &
y_t = -{\alpha \over 2} \lambda^{-1} e^{U}
      +{\alpha \over 2} \lambda^{-1} e^{-U} y^2.
\label{eqRiccatiZt}
\end{eqnarray}
The second-order matrix Lax pair 
results from the linearization $y=\psi_1 / \psi_2$
\begin{eqnarray}
& &
(\partial_x - L) \pmatrix{\psi_1 \cr \psi_2 \cr}=0,\
L=
\pmatrix{U_x/2 & \lambda \cr \lambda & - U_x/2 \cr},\
\\
& &
(\partial_t - M) \pmatrix{\psi_1 \cr \psi_2 \cr}=0,\
M=
\pmatrix{0 & - (\alpha/2) \lambda^{-1} e^{U} \cr 
             - (\alpha/2) \lambda^{-1} e^{- U} & 0 \cr}.
\end{eqnarray}
The auto-BT 
results from the substitution
$y= e^{-(u - U)/2}$ into (\ref{eqRiccatiZx})--(\ref{eqRiccatiZt})
\begin{eqnarray}
& &
(u+U)_x = - 4 \lambda \sinh {u - U \over 2},\
\\
& &
(u-U)_t = 2 \alpha \lambda^{-1} \sinh {u + U \over 2}.
\end{eqnarray}
The ODE part of the BT is a Riccati equation,

% ==========================================================================
\section{Truncation results}
\indent

% *********************************************************** J'EN SUIS LA

Table \ref{table} summarizes,
for a sample of PDEs,
the currently best method to obtain its Lax pair, Darboux and B\"acklund
transformations from a truncation.
The reference given in each entry is the place where the right method has been
applied, and earlier references may be found in it.

The ``?'' in the AKNS system entry 
(the one whose NLS is a reduction)
means that the method has not yet been applied to it.
Nevertheless, its BT can be found by applying the one-family method
followed by four involutions \cite{CMGalli95}.

\tabcolsep=1.5truemm
\tabcolsep=0.5truemm

\begin{table}[h] % [p]
\caption[garbage]{
The relevant truncation for some well known $1+1$-dimensional PDEs.
The successive columns are~:
the usual name of the PDE
(a p means the potential equation),
its number of families
(a * indicates that only one family is relevant, see details in Ref),
the order of its Lax pair,
the truncation variable 
(notation $Y_1=\psi_x / \psi, Y_2=\psi_{xx} / \psi$),
the link between $\tau$ and $\psi$,
the singularity orders of $u$ and $E(u)$,
the Fuchs indices (without the ever present $-1$),
the number of determining equations,
the reference.
}
\vspace{0.2truecm}
\begin{center}
\begin{tabular}{| l | l | l | l | l | l | l | l | l |}
\hline % \hline % ********************************************************
\hline % \hline % ********************************************************
Name
&
f
&
Lax
&
Trunc.~var.
&
$\tau$
&
$-p:-q$
&
indices
&
nb. det. eq.
&
Ref
\\ \hline % \hline % ********************************************************
KdV     % Name
&
$1$     % Nb of families
&
$2$     % Scattering order
&
$\chi$  % Truncation variable
&
$\psi$  % Link between $\tau$ and $\psi$
&
$2:5$   % -p:-q
&
$4,6$   % Fuchs indices
&
$2$     % Number of determining equations 3, 5
&
\cite{WTC} % Ref
\\ \hline % \hline % ********************************************************
p-mKdV     % Name
&
$2$     % Nb of families
&
$2$     % Scattering order
&
$Y$     % Truncation variable
&
$\psi$  % Link between $\tau$ and $\psi$
&
$0:3$   % -p:-q
&
$0,4$   % Fuchs indices
&
$3$     % Number of determining equations 1, 3, 5
&
\cite{Pickering1996} % Ref
\\ \hline % \hline % ********************************************************
sine-Gordon    % Name
&
$2$     % Nb of families
&
$2$     % Scattering order
&
$Y$     % Truncation variable
&
$\psi$  % Link between $\tau$ and $\psi$
&
$0:2$   % -p:-q
&
$2$     % Fuchs indices
&
$2$     % Number of determining equations 1,3
&
\cite{Pickering1996} % Ref
\\ \hline % \hline % ********************************************************
Broer-Kaup     % Name
&
$2$     % Nb of families
&
$2$     % Scattering order
&
$Y$     % Truncation variable
&
$\psi$  % Link between $\tau$ and $\psi$
&
$0:4$   % -p:-q
&
$0,3,4$ % Fuchs indices
&
$4$     % Number of determining equations 1, 2, 6, 7
&
\cite{Pickering1996} % Ref
\\ \hline % \hline % ********************************************************
pp-Boussinesq    % Name
&
$1$     % Nb of families
&
$3$     % Scattering order
&
$\chi$  % Truncation variable
&
$\psi$  % Link between $\tau$ and $\psi$
&
$0:4$   % -p:-q
&
$0,1,6$ % Fuchs indices
&
$6$     % Number of determining equations
&
\cite{MCGallipoli1991} % Ref
\\ \hline % \hline % ********************************************************
p-SK    % Name
&
$1*$     % Nb of families
&
$3$     % Scattering order
&
$(Y_1,Y_2)$  % Truncation variable
&
$\psi$  % Link between $\tau$ and $\psi$
&
$1:6$   % -p:-q
&
$1,2,3,10$   % Fuchs indices
&
$6$     % Number of determining equations
&
\cite{MC1998} % Ref
\\ \hline % \hline % ********************************************************
p-KK    % Name
&
$1*$     % Nb of families
&
$3$     % Scattering order
&
$(Y_1,Y_2)$  % Truncation variable
&
G25($\psi$) % Link between $\tau$ and $\psi$
&
$1:6$   % -p:-q
&
$1,3,5,7$   % Fuchs indices
&
$38$     % Number of determining equations
&
\cite{MC1998} % Ref
\\ \hline % \hline % ********************************************************
Tzitz\'eica % Name
&
$1*$     % Nb of families
&
$3$     % Scattering order
&
$(Y_1,Y_2)$  % Truncation variable
&
$\psi$  % Link between $\tau$ and $\psi$
&
$2:6$   % -p:-q
&
$2$   % Fuchs indices
&
$10$     % Number of determining equations
&
\cite{CMG1998} % Ref
\\ \hline % \hline % ********************************************************
AKNS system % Name
&
$4$     % Nb of families
&
$2$     % Scattering order
&
$?$  % Truncation variable
&
$ $  % Link between $\tau$ and $\psi$ ?
&
$1:3,1:3$   % -p:-q
&
$0,3,4$   % Fuchs indices
&
$ $     % Number of determining equations
&
\cite{CMGalli95} % Ref
\\ \hline % \hline % ********************************************************
\end{tabular}
\end{center}
\label{table}
\end{table}

\vfill \eject

% ==========================================================================
\section{Conclusion}
\indent

The truncation procedure is not purely academic,
as could be thought from all the above integrable examples.
There exist many challenging problems,
in particular in nonlinear optics and spatiotemporal intermittency 
\cite{BN1985,vanHecke},
in which the equations, although nonintegrable, possess some regular
``patterns'' which could well be described by exact particular solutions.
The difficulty to find them \cite{CM1993} comes from the good guess which
must be made for the entire functions $\psi$,
which do not necessarily satisfy a linear system any more.

% ==========================================================================
\section*{Acknowledgments}
\indent

The author would like to thank 
the Bharatidasan University for its warm hospitality and support,
and the Minist\`ere des affaires \'etrang\`eres for travel support.

\end{document}